\documentstyle[aps,preprint,epsfig]{revtex}
\begin{document}
\tighten
\draft
\preprint{IFA-96/40}
\title{Unitarity, Chiral Perturbation Theory, and Meson Form Factors}
\author{Torben Hannah\thanks{Electronic address: hannah@dfi.aau.dk}}
\address{Institute of Physics and Astronomy, Aarhus University,
DK-8000 Aarhus C, Denmark}
\maketitle
\begin{abstract}
The inverse-amplitude method is applied to the one-loop chiral
expansion of the pion, kaon, and $K_{l3}$ form factors. Since these
form factors are determined by the same chiral low-energy constants,
it is possible to obtain finite predictions for the inverse-amplitude
method. It is shown that this method clearly improves one-loop chiral
perturbation theory, and a very good agreement between
the inverse-amplitude method and the experimental information is
obtained. This suggests that the inverse-amplitude method is a rather
systematic way of improving chiral perturbation theory.
\end{abstract}
\pacs{PACS: 13.40.Gp, 11.30.Rd, 11.55.Fv, 13.20.Eb}

\section{INTRODUCTION}
\label{sec:intro}

Chiral perturbation theory (ChPT) \cite{ref:We79,ref:GL84} is a
rigorous methodology for QCD in the low-energy region. With this
methodology, one obtains an expansion in terms of the energy or,
equivalently, in the number of loops. However, to go beyond leading
order, one has to introduce a number of phenomenological low-energy
constants. At next-to-leading order in the chiral expansion, there is
only a small number of low-energy constants, which are known rather
accurately \cite{ref:GL84,ref:BCG94}. Therefore, ChPT to one loop
provides finite predictions for many different processes, which in
general agree well with the experimental low-energy data.

Nevertheless, it is important to have an estimate for the higher-order
corrections in the chiral expansion. Of course, the leading correction
to the one-loop result could be found by an explicit two-loop
calculation in the framework of ChPT \cite{ref:BGS94}. Unfortunately,
this will introduce further phenomenological low-energy constants,
which have to be fixed before finite predictions are possible.
Moreover, in the presence of resonances, even the two-loop result will
only be applicable well below the resonance energy. Alternatively, in
order to estimate the higher-order corrections, one could combine
unitarity and dispersion relations with one-loop ChPT. A rather
general way to make this combination is by the inverse-amplitude
method as originally discussed by Truong and collaborators
\cite{ref:Tru88,ref:DHT90}.

This method has previously been applied separately to both the pion
\cite{ref:Tru88,ref:GM90} and the $K_{l3}$ \cite{ref:BT95} form
factors. However, these form factors together with the charge kaon
form factor are determined by the same chiral low-energy constants
\cite{ref:GL85}. Therefore, the inverse-amplitude method can be
applied simultaneously to the pion, kaon, and $K_{l3}$ form factors in
order to obtain finite predictions. In this paper, these predictions
are compared with one-loop ChPT and the available experimental
information in order to establish whether the inverse-amplitude method
is a systematic way of improving ChPT. In Sec. \ref{sec:method},
an overview of the inverse-amplitude method as applied to the meson form
factors is given. The comparison with the experimental information is
discussed in Sec. \ref{sec:compare}, and the conclusion of this work
is given in Sec. \ref{sec:con}.

\section{INVERSE-AMPLITUDE METHOD}
\label{sec:method}

The inverse-amplitude method was originally applied to the pion vector
($F_V^{\pi}$) and scalar ($F_S^{\pi}$) form factors \cite{ref:Tru88}.
This method can also be applied to the $K_{l3}$ vector ($f_+^{K\pi}$)
and scalar ($f_0^{K\pi}$) form factors, which describe the decays
$K\rightarrow\pi l\nu$, where $l=e,\mu$. All these form factors depend
on a single kinematical variable $t$, giving by the square of the
four-momentum transfer. They are analytical functions in the complex
$t$ plane with cuts\footnote{In the following, the notation
$\alpha$ and $\beta = \pi$ or $K$ is used in order to describe both
the pion and the $K_{l3}$ form factors with the same formulas.
Furthermore, the generic symbol $F$ is used for
any one of these form factors.}
starting at the $\pi\pi$ or $\pi K$ threshold
$t=(M_{\alpha}+M_{\beta})^2$.
In the elastic region, one has the unitarity relation
\begin{equation}
\label{eq:funi}
{\rm Im}F(t) = \sigma (t)F^{\ast}(t)t^I_l(t) ,
\end{equation}
where $\sigma$ is the phase space factor and $t^I_l$ denotes the
relevant $\pi\pi$ or $\pi K$ partial wave. This relation, in
practice, is useful up to the $K\bar{K}$ threshold for the pion form
factors and up to the $K\eta$ threshold for the $K_{l3}$ form factors.

The pion and $K_{l3}$ form factors have been determined by Gasser and
Leutwyler to one loop in the chiral expansion \cite{ref:GL85}. The
result for these form factors may be written as
\begin{equation}
\label{eq:ChPTform}
F(t) = F^{(0)}(t)+F^{(1)}(t) ,
\end{equation}
where $F^{(0)}$ denotes the leading order contribution and $F^{(1)}$ is
the one-loop correction. Since the normalized form factor
$F_S^{\pi}(0)=1$ is used in the following, one has in all cases
$F^{(0)}=1$. The one-loop result satisfies unitarity perturbatively:
\begin{equation}
\label{eq:puni}
{\rm Im}F^{(1)}(t) = \sigma (t)F^{(0)}(t)t^{I(0)}_l(t) = \sigma (t)
t^{I(0)}_l(t) ,
\end{equation}
where $t^{I(0)}_l$ is the relevant leading order $\pi\pi$ or
$\pi K$ partial wave. This relation holds in one-loop ChPT up to
$t=(M_{\gamma}+M_{\delta})^2$, where $\gamma$ and $\delta = K$ or
$\eta$, i.e., up to the $K\bar{K}$ or $K\eta$ threshold. Because of the
analytical structure of the form factors, the one-loop ChPT result
(\ref{eq:ChPTform}) may be expressed in form of a twice-subtracted
dispersion relation. The subtraction constants are given by expanding
the one-loop result to $O(t)$ as $F^{(0)}=1$ and
$F^{(1)}=a_0+a_1t$. Furthermore, in the region
$(M_{\alpha}+M_{\beta})^2\leq t\leq (M_{\gamma}+M_{\delta})^2$ the
perturbative unitarity relation (\ref{eq:puni}) may be used. Thus, the
one-loop ChPT result (\ref{eq:ChPTform}) may be written as
\begin{eqnarray}
\label{eq:ChPTdisp}
F^{(0)}(t) & = & 1 ,\nonumber \\
F^{(1)}(t) & = & a_0+a_1t+\frac{t^2}{\pi}
\int_{(M_{\alpha}+M_{\beta})^2}^{(M_{\gamma}+M_{\delta})^2}
\frac{\sigma (t')t^{I(0)}_l(t')dt'}{t'^2(t'-t-i\epsilon )}
\nonumber \\
&& +\frac{t^2}{\pi}\int_{(M_{\gamma}+M_{\delta})^2}^{\infty}
\frac{{\rm Im}F^{(1)}(t')dt'}{t'^2(t'-t-i\epsilon )} .
\end{eqnarray}
In this dispersion relation unitarity was only applied
perturbatively. Therefore, it could well be that there are other more
useful ways to use one-loop ChPT than the truncation of the chiral
expansion (\ref{eq:ChPTform}). In order to improve ChPT, one could use
exact unitarity and dispersion relations together with ChPT, since the
combination of these two approaches is likely to be more powerful
than either one separately.

A rather general way to make this combination is by the
inverse-amplitude method \cite{ref:Tru88,ref:DHT90}. This approach
consists in writing down a dispersion relation for the inverse of the
form factor $1/F$. In the elastic region, the use of exact unitarity
(\ref{eq:funi}) gives ${\rm Im}(1/F)=-{\rm Im}F/|F|^2=
-\sigma t^I_l/F=-\sigma t^{I(0)}_l$ to one-loop order in the chiral
expansion. Above this region, the approximation ${\rm Im}(1/F)=
-{\rm Im}F^{(1)}$ may be used to the same order. Finally, the
subtraction constants are given to one-loop order in ChPT as
$1-a_0-a_1t$. Thus, neglecting any pole contribution arising from
possible zeros in the form factors \cite{ref:Tru88,ref:DHT90}, the
dispersion relation for the inverse form factor becomes
\begin{eqnarray}
\label{eq:Indisp}
\frac{1}{F(t)} & = & 1-a_0-a_1t-\frac{t^2}{\pi}
\int_{(M_{\alpha}+M_{\beta})^2}^{(M_{\gamma}+M_{\delta})^2}
\frac{\sigma (t')t^{I(0)}_l(t')dt'}{t'^2(t'-t-i\epsilon )}
\nonumber \\
&& -\frac{t^2}{\pi}\int_{(M_{\gamma}+M_{\delta})^2}^{\infty}
\frac{{\rm Im}F^{(1)}(t')dt'}{t'^2(t'-t-i\epsilon )} .
\end{eqnarray}
Comparing this dispersion relation with the corresponding one for
one-loop ChPT, Eqs. (\ref{eq:ChPTdisp}), one finds that the form
factor may be written as
\begin{equation}
\label{eq:Inform}
F(t) = \frac{1}{1-F^{(1)}(t)} .
\end{equation}
This is formally equivalent to the [0,1] Pad\'{e} approximant applied
to one-loop ChPT. It will satisfy the unitarity relation
(\ref{eq:funi}) exactly provided that the $t^I_l$ partial waves are
given by
\begin{equation}
\label{eq:partial}
t^I_l(t) = \frac{t^{I(0)}_l(t)}{1-F^{(1)}(t)} .
\end{equation}
This result satisfies the unitarity relation for the $\pi\pi$ and
$\pi K$ partial waves, and so the inverse-amplitude method is a
consistent approach. However, this method can also be applied directly
to the $\pi\pi$ and $\pi K$ partial waves \cite{ref:DHT90} by means of
which one obtains a slightly different result than the one in
Eq. (\ref{eq:partial}). Still, the numerical difference between these
two results for $t^I_l$ is small due to the fact that the approximation
$t^I_l/F\simeq t^{I(0)}_l$ is well satisfied \cite{ref:Tru88}.
The reason for expecting that the inverse-amplitude method is superior
to the truncation of the chiral expansion is indeed based upon the
fact that this approximation is better satisfied than the chiral
approximation $F^{\ast}t^I_l\simeq t^{I(0)}_l$.

The inverse-amplitude method can also be applied to the one-loop ChPT
result for the charge kaon
vector form factor $F_V^{K^+}$ \cite{ref:GL85} with a final result
similar to Eq. (\ref{eq:Inform}). In this case, unitarity relates
${\rm Im}F_V^{K^+}$ to $F_V^{\pi}$ and the $\pi\pi\rightarrow K\bar{K}$
partial wave $t^1_1$. This unitarity relation implies that the
approximation ${F_V^{\pi}}^{\ast}t^1_1/|F_V^{K^+}|^2\simeq t^{1(0)}_1$
is used in the elastic region for the inverse-amplitude method,
whereas one-loop ChPT is based upon the approximation
${F_V^{\pi}}^{\ast}t^1_1\simeq t^{1(0)}_1$. Since the former
approximation is better satisfied than the latter, one would also in
this case expect that the inverse-amplitude method is superior to the
truncation of the chiral expansion. Finally, it should be noticed that
the inverse-amplitude method cannot be applied to the neutral kaon
vector form factor $F_V^{K^0}$ \cite{ref:GL85}. This is due to the
fact that the leading order result for $F_V^{K^0}$ vanishes.
Therefore, this form factor will not be considered in the following.

\section{COMPARISON WITH EXPERIMENT}
\label{sec:compare}

The meson form factors depend on the pion decay constant $F_{\pi}$ and
the meson masses $M_{\pi}$, $M_K$, and $M_{\eta}$. Unless stated
otherwise, the values $F_{\pi}=92.4$ MeV, $M_{\pi}=139.6$ MeV,
$M_K=493.7$ MeV, and $M_{\eta}=548.8$ MeV are used in the following.
In addition, the renormalization scale is
set to $\mu =M_{\eta}$. Having fixed this,
the vector form factors $F_V^{\pi}$, $F_V^{K^+}$, and $f_+^{K\pi}$ are
determined entirely by the chiral low-energy constant $L^r_9$. The
most reliable way to pin down this low-energy constant is to consider
the slope of the vector form factors at $t=0$, since this should
minimize the effect of the higher-order corrections. From these
slopes, the best determination of $L^r_9$ is obtained with the rather
precise experimental value of the pion charge radius
$\langle r^2\rangle_V^{\pi}=0.439\pm 0.008$ ${\rm fm}^2$
\cite{ref:NA786}. This gives the value
$L^r_9=(7.31\pm 0.15)\times 10^{-3}$ for both the inverse-amplitude
method, Eq. (\ref{eq:Inform}), and one-loop ChPT,
Eq. (\ref{eq:ChPTform}).

The scalar form factors $F_S^{\pi}$ and $f_0^{K\pi}$
depend on the chiral low-energy constant $L^r_5$. This low-energy
constant can be determined in an independent way from the ratio
$F_K/F_{\pi}$ \cite{ref:GL84}. With the experimental values for
$F_{\pi}$ and $F_K$ given in \cite{ref:PDG94}, one obtains
$L^r_5=(2.26\pm 0.14)\times 10^{-3}$. Finally, $F_S^{\pi}$ also
depends on the low-energy constant $L^r_4$, which has been
determined using large $N_c$ arguments \cite{ref:GL84}. To
recapitulate, the values
\begin{eqnarray}
\label{eq:L954}
L^r_9(M_{\eta}) & = & (7.31\pm 0.15)\times 10^{-3}, \nonumber \\
L^r_5(M_{\eta}) & = & (2.26\pm 0.14)\times 10^{-3}, \nonumber \\
L^r_4(M_{\eta}) & = & (0.0\pm 0.5)\times 10^{-3}
\end{eqnarray}
are used in the following both for the inverse-amplitude method,
Eq. (\ref{eq:Inform}), and in the case of one-loop ChPT,
Eq. (\ref{eq:ChPTform}). In the figures, only the central values of
these low-energy constants are used.

\subsection{Pion and kaon form factors}
\label{subsec:pKform}

Having fixed the low-energy constants, it is possible to obtain finite
predictions. Considering first the pion and kaon form factors, these
may be expanded around $t=0$ as
\begin{equation}
\label{eq:pKex}
F(t) = 1+\mbox{$\frac{1}{6}$}\langle r^2\rangle t+ct^2+\cdots .
\end{equation} 
In Table \ref{Table1}, the predictions for $\langle r^2\rangle$ and
$c$ obtained from one-loop ChPT and the inverse-amplitude method are
compared with the experimental information.
Since the experimental value
of $\langle r^2\rangle_V^{\pi}$ was used in order to determine the
low-energy constant $L^r_9$, there are of course no predictions in
this case. For $c_V^{\pi}$, the prediction obtained from the
inverse-amplitude method agrees very well with the experimental
information, whereas the prediction from one-loop ChPT is far too
small. The experimental range for $c_V^{\pi}$ was obtained in Ref.
\cite{ref:GM90} from two different parametrizations of $F_V^{\pi}$
\cite{ref:Bar85,ref:GS68} together with the experimental value of
$\langle r^2\rangle_V^{\pi}$.
With regard to $\langle r^2\rangle_S^{\pi}$,
the predictions agree reasonably well with the experimental
information. In the case of $c_S^{\pi}$ this is only true for the
inverse-amplitude method, whereas the prediction from one-loop ChPT is
too small. The experimental range for $\langle r^2\rangle_S^{\pi}$ and
$c_S^{\pi}$ was obtained in Ref. \cite{ref:GM90} from a dispersive
analysis of $F_S^{\pi}$ \cite{ref:DGL90}. Finally, the predictions for
$\langle r^2\rangle_V^{K^+}$ seem to be somewhat larger than the
central experimental value \cite{ref:Amen86}. However,
$\langle r^2\rangle_V^{K^+}$ is related to the slopes of the other
meson vector form factors via a low-energy theorem \cite{ref:GL85}.
With the experimental values of these other slopes, one obtains
$\langle r^2\rangle_V^{K^+}=0.40\pm 0.07$ ${\rm fm}^2$, which is in
excellent agreement with the predictions.

\begin{figure}
\centering{\epsfig{file=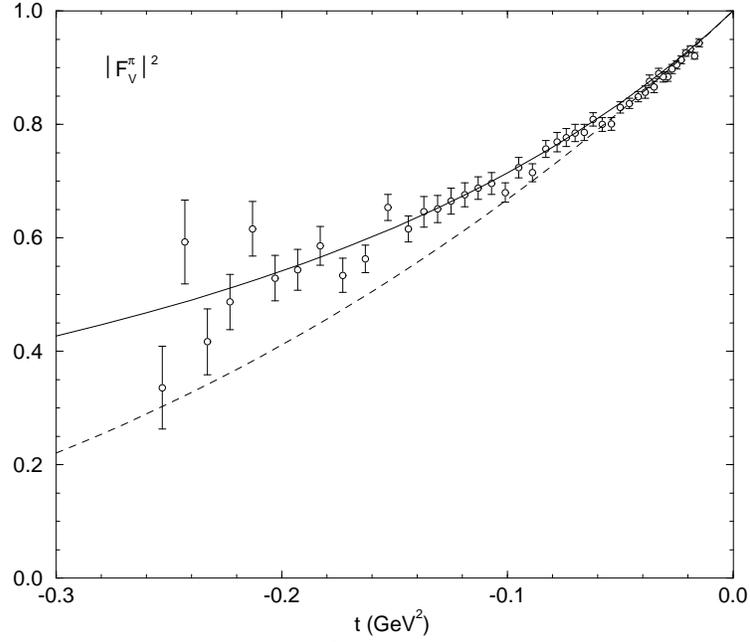,height=12cm,angle=-90}}
\caption{The pion vector form factor $|F_V^{\pi}|^2$ in the spacelike
region. The solid line is the inverse-amplitude method, the dashed
line one-loop ChPT, and the experimental data are from Ref.
\protect\cite{ref:NA786}.}
\label{Fig1}
\end{figure}

The pion vector form factor $F_V^{\pi}$ is well known experimentally
both in the spacelike region $t<0$ \cite{ref:NA786} and in the
timelike region $t>4M_{\pi}^2$ \cite{ref:Bar85}. In Fig. \ref{Fig1},
the predictions for $|F_V^{\pi}|^2$
are compared with the spacelike experimental data \cite{ref:NA786}.
This shows that the inverse-amplitude method agrees very well with the
experimental data over the whole $t$ region displayed, whereas
one-loop ChPT only describes the data accurately near $t=0$.
Instead of having fixed $L^r_9$ from the experimental value of
$\langle r^2\rangle_V^{\pi}$, this low-energy constant could also
be determined by a fit to the spacelike experimental data
\cite{ref:BC88}. With this approach, one also
finds that the inverse-amplitude method improves one-loop ChPT, in
that the former gives a significantly smaller value of
$\chi^2/N_{\rm DF}$ than the latter. Furthermore, the value of $L^r_9$
obtained from fitting the inverse-amplitude method is much more
consistent with Eqs. (\ref{eq:L954})
than the corresponding value obtained from one-loop ChPT. Of course,
reducing the energy range in these fits improves the value of
$\chi^2/N_{\rm DF}$ for one-loop ChPT, but this value will still be
somewhat larger than the corresponding one for the inverse-amplitude
method.

\begin{figure}
\centering{\epsfig{file=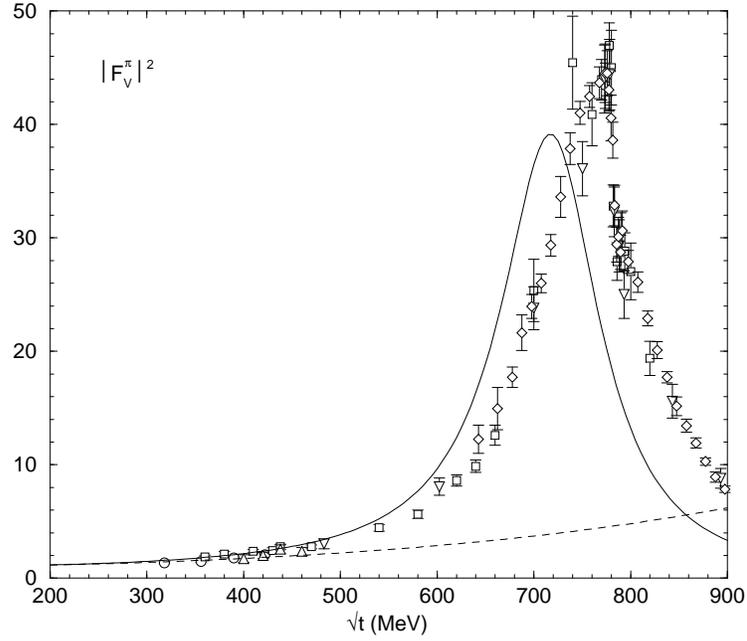,height=12cm,angle=-90}}
\caption{The pion vector form factor $|F_V^{\pi}|^2$ in the timelike
region. The solid line is the inverse-amplitude method, the dashed
line one-loop ChPT, and the experimental data are from Ref.
\protect\cite{ref:Amen84} (circles), Refs.
\protect\cite{ref:Bar85,ref:Ani83} (squares), Ref.
\protect\cite{ref:Vas81} (triangles), Ref. \protect\cite{ref:Quen78}
(inverted triangles), and Ref. \protect\cite{ref:Kur84} (diamonds).}
\label{Fig2}
\end{figure}

In Fig. \ref{Fig2}, the predictions for $|F_V^{\pi}|^2$
are compared with the timelike experimental data
\cite{ref:Bar85,ref:Amen84,ref:Ani83,ref:Vas81,ref:Quen78,ref:Kur84}.
Because of the $\rho$(770) vector meson,
these data show a clear resonance
behavior. This behavior is also obtained from the inverse-amplitude
method, whereas one-loop ChPT only accounts for the low-energy tail of
this resonance. With $L^r_9$ given by Eqs. (\ref{eq:L954}), the
inverse-amplitude method gives a resonance\footnote{Resonances are
defined to be where the phase passed ${\pi}/2$.} in the range 723--738
MeV, whereas a resonance at 770 MeV implies $L^r_9=6.57\times
10^{-3}$. However, with this value of $L^r_9$ the inverse-amplitude
method does not account accurately for either the height of the
resonance or the rather precise experimental data in the spacelike
region. Furthermore, this value of $L^r_9$ is rather inconsistent
with the experimental values of both $\langle r^2\rangle_V^{\pi}$
and the slope of $f_+^{K\pi}$.
Hence, the inverse-amplitude method applied on one-loop ChPT only
approximates the behavior of the form factor in the resonance region.
Nevertheless, if the inverse-amplitude method was applied on two-loop
ChPT \cite{ref:GM90}, it is likely that the agreement could be
improved in the resonance region.

\begin{figure}
\centering{\epsfig{file=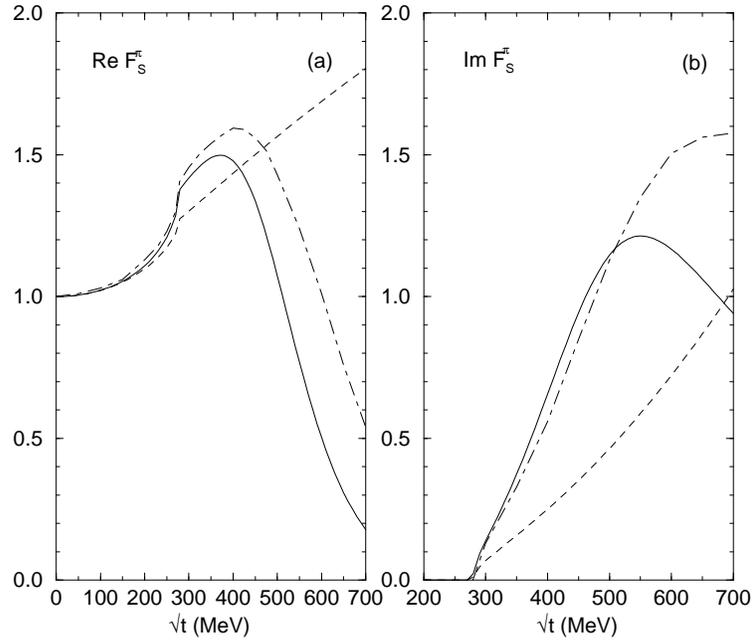,height=12cm,angle=-90}}
\caption{The real (a) and imaginary (b) part of the pion scalar form
factor $F_S^{\pi}$. The solid line is the inverse-amplitude method,
the dashed line one-loop ChPT, and the dashed-dotted line solution B
from a dispersive analysis \protect\cite{ref:GM90,ref:DGL90}.}
\label{Fig3}
\end{figure}

Considering the pion scalar form factor $F_S^{\pi}$, this is not
directly accessible to experiment. However, it has been determined in
terms of the experimental phase shifts for the $\pi\pi/K\bar{K}$
system by a dispersive analysis \cite{ref:GM90,ref:DGL90}. In Fig.
\ref{Fig3}, the predictions are compared
with one of the solutions (B) from this
dispersive analysis. It is observed that the inverse-amplitude method
also in this case improves one-loop ChPT, and that the latter only
agrees with the dispersive analysis below the $\pi\pi$ threshold.
This is due to the strong final-state interaction, which makes the
higher-order corrections important even at low energies.

\begin{figure}
\centering{\epsfig{file=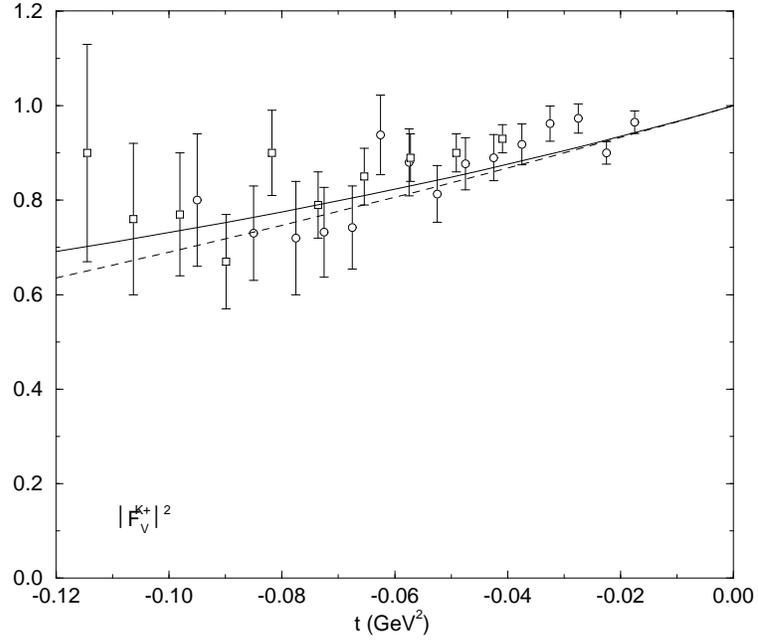,height=12cm,angle=-90}}
\caption{The charge kaon vector form factor $|F_V^{K^+}|^2$ in the
spacelike region. The solid line is the inverse-amplitude method, the
dashed line one-loop ChPT, and the experimental data are from Ref.
\protect\cite{ref:Amen86} (circles) and Ref. \protect\cite{ref:Dal80}
(squares).}
\label{Fig4}
\end{figure}

Finally, in Fig. \ref{Fig4} the predictions for $|F_V^{K^+}|^2$
are compared with the spacelike
experimental data \cite{ref:Amen86,ref:Dal80}. The two theoretical
approaches give rather similar results in the energy region displayed,
and both agree rather well with the not very conclusive experimental
data. At low energies, however, the main part of the data seems to be
systematically above the theoretical predictions. This implies a
smaller experimental value of $\langle r^2\rangle_V^{K^+}$ than
obtained theoretically. However, as already discussed, the
experimental data for the other meson vector form factors support the
value of $\langle r^2\rangle_V^{K^+}$ obtained theoretically.

\subsection{$K_{l3}$ form factors}
\label{subsec:Kl3form}

Turning to the $K_{l3}$ form factors, the notation $K_{l3}^+$ and
$K_{l3}^0$ is used for the decays $K^{\pm}\rightarrow\pi^0l^{\pm}\nu$
and $K^0\rightarrow\pi^{\pm}l^{\mp}\nu$, respectively. Hence, the
values $M_{\pi}=M_{\pi^0}=135.0$ MeV and $M_K=M_{K^+}=493.7$ MeV are
used in the case of the $K_{\l3}^+$ form factors, whereas
$M_{\pi}=M_{\pi^+}=139.6$ MeV and $M_K=M_{K^0}=497.7$ MeV are used for
the $K_{\l3}^0$ form factors.

Analysis of the $K_{l3}$ data frequently assumes a linear dependence
of $f_+^{K\pi}$ and $f_0^{K\pi}$ on $t$, i.e., only the first two terms
in the expansion
\begin{equation}
\label{eq:Kl3ex}
f_{+,0}^{K\pi}(t)=f_{+,0}^{K\pi}(0)\left[
1+\lambda_{+,0}\frac{t}{M_{\pi}^2}+\cdots \right] .
\end{equation}
In Table \ref{Table2}, the predictions for $\lambda$ obtained
from one-loop ChPT and the inverse-amplitude method are compared with
the experimental data. For the slope $\lambda_+$, the most precise
experimental determination is obtained via the decays $K_{e3}^+$ and
$K_{e3}^0$ \cite{ref:PDG94}. Hence, these experimental results are
given in Table \ref{Table2}. From the other $K_{l3}$ decays
$K_{\mu 3}^+$ and $K_{\mu 3}^0$, one obtains more uncertain values for
$\lambda_+$ which are, however, still consistent with the results
displayed in Table \ref{Table2}. From this table, it is observed that
for $\lambda_+$ both one-loop ChPT and the inverse-amplitude method
agree remarkably well with the experimental values. Nevertheless, the
inverse-amplitude method seems to agree slightly better with the
central experimental data than one-loop ChPT.

With regard to $\lambda_0$, the experimental situation is rather
unsatisfactory. This slope is only accessible experimentally via the
$K_{\mu 3}$ decays, where the result of Donaldson {\em et al.}
\cite{ref:Don74}, $\lambda_0=0.019\pm 0.004$, dominates the statistics
in the $K_{\mu 3}^0$ case. This experiment also measured $\lambda_+$
with the result $0.030\pm 0.003$ in agreement with 
the $K_{e3}^0$ value. However, more recent
$K_{\mu 3}^0$ experiments find a substantially larger value for
$\lambda_0$ \cite{ref:PDG94}. The experimental situation is not
more satisfactory for the $K_{\mu 3}^+$ decay which also in
general have poor statistics \cite{ref:PDG94}. In Table
\ref{Table2} only the high statistics measurement of $\lambda_0$ by
Donaldson {\em et al.} is displayed. It is observed that both one-loop
ChPT and the inverse-amplitude method agree well with this
result. With regard to the more recent larger values of $\lambda_0$
as, for instance, given by the measurement $\lambda_0=0.0341\pm 0.0067$
\cite{ref:Bir81}, this result
is clearly inconsistent with both one-loop
ChPT and the inverse-amplitude method. Forthcoming kaon facilities
like DA$\Phi$NE \cite{ref:DAPHNE95} could very well settle the issue
concerning the experimental value of $\lambda_0$, and thereby test the
predictions in much greater detail.

\begin{figure}
\centering{\epsfig{file=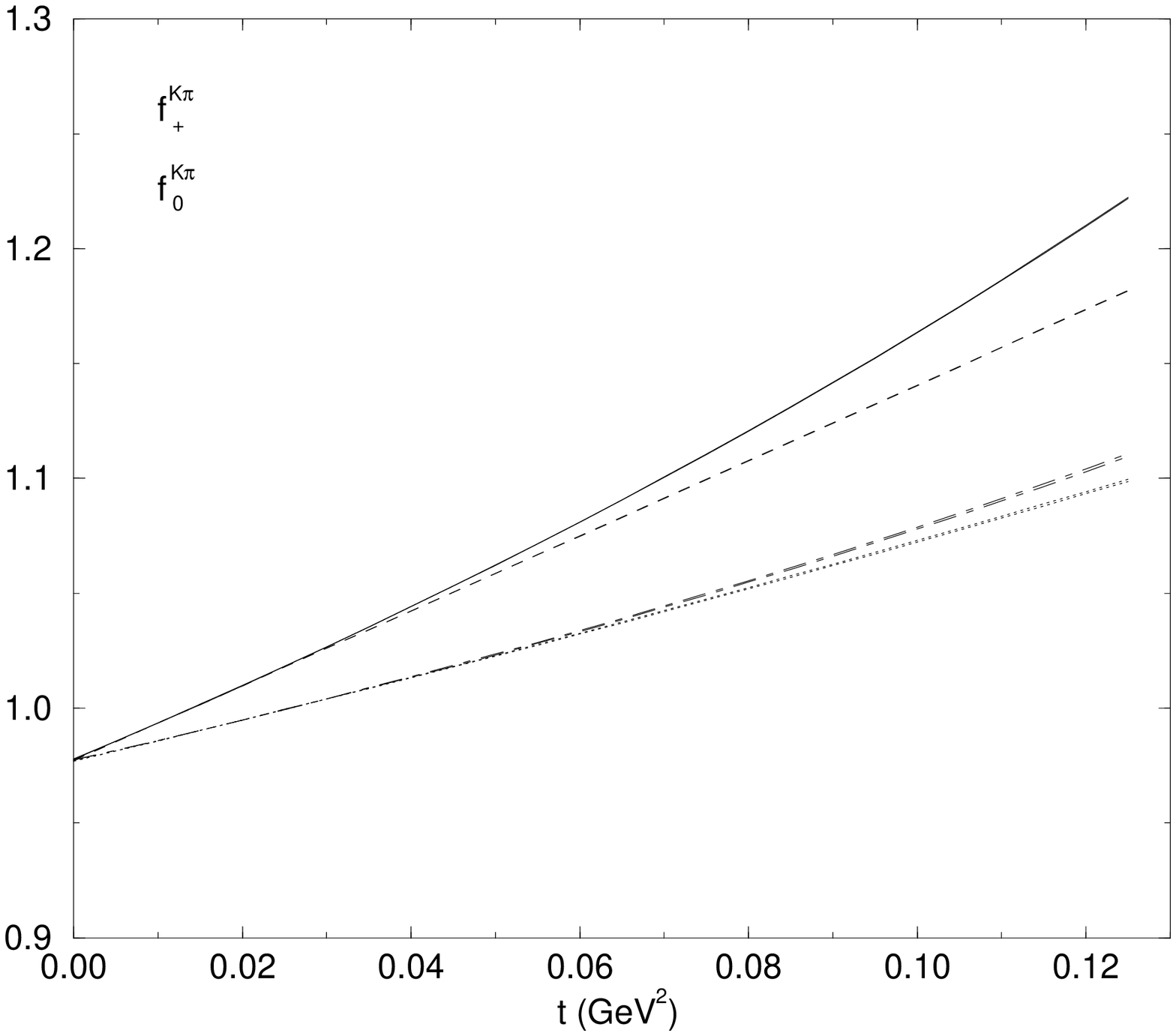,height=12cm,angle=-90}}
\caption{The $K_{l3}$ form factors in the physical region. The
numerical difference between the $K_{l3}^+$ and $K_{l3}^0$ form
factors is negligible in this region. The solid line
(inverse-amplitude method) and dashed line (one-loop ChPT) are for the
$K_{l3}$ vector form factor $f_+^{K\pi}$, whereas the dash-dotted
line (inverse-amplitude method) and dotted line (one-loop ChPT) are
for the scalar form factor $f_0^{K\pi}$.}
\label{Fig5}
\end{figure}

Finally, Fig. \ref{Fig5} shows the predictions for the $K_{l3}$ form
factors in the physical region. It is observed
that the inverse-amplitude
method only modifies one-loop ChPT modestly in this region.
At higher energies, however, the difference becomes more pronounced.
Furthermore, this figure shows that
the inverse-amplitude method deviates
more clearly from linearity for $f_+^{K\pi}$ than in the case of
$f_0^{K\pi}$, which should also
be expected due to the presence of the
$K^*$(892) vector resonance. Indeed, the inverse-amplitude method
generates a resonance for $f_+^{K\pi}$ in the range 761--777 MeV,
whereas a resonance at 892 MeV implies $L^r_9=5.33\times 10^{-3}$.
However, this value of $L^r_9$ is clearly inconsistent with the rather
precise experimental information on both
the $K_{l3}$ and the pion vector
form factors. Thus, the inverse-amplitude method applied on one-loop
ChPT only accounts qualitatively for the $K^*$(892) resonance.

As already mentioned, the experimental analysis of the $K_{l3}$ data
usually assumes a linear dependence of the form factors in
the physical region. Thus, in order to test this assumption, it is
important to estimate the higher-order corrections in the chiral
expansion. With the inverse-amplitude method one finds that this
assumption is well satisfied for the $K_{l3}$ scalar form factor
$f_0^{K\pi}$. Therefore, the experimental uncertainty in the
determination of $\lambda_0$ is not likely to be due to the assumption
of linearity in the analysis of the data. For the $K_{l3}$ vector form
factor $f_+^{K\pi}$, one finds that the assumption of linearity is
less satisfactory. Therefore, in the experimental search for scalar
and tensor couplings in the $K_{l3}$ decays, these nonlinearities
should be included in the analysis of the experimental data. The most
recent experimental measurement points towards the
conclusion that the presence of scalar and tensor couplings or
nonlinearities in the $f_+^{K\pi}$ form factor cannot be excluded
\cite{ref:Aki91}. However, this conclusion has to await verification
from forthcoming kaon facilities such as DA$\Phi$NE \cite{ref:DAPHNE95}.

\section{CONCLUSION}
\label{sec:con}

The inverse-amplitude method is based upon the use of unitarity and
dispersion relations together with ChPT. Therefore, it is expected
that this method improves the truncation of the chiral expansion.
However, to establish this more firmly, it is important to obtain
finite predictions for the inverse-amplitude method, which can be
compared with the corresponding predictions from ChPT.

This is possible for the pion, kaon, and $K_{l3}$ form factors, since
these are determined by the same chiral low-energy constants.
Therefore, the inverse-amplitude method has been applied
simultaneously to the one-loop chiral expansion of these form factors.
After having determined the chiral low-energy constants, the finite
predictions are presented. A comparison with the experimental
information shows that the inverse-amplitude method agrees
significantly better with the data than one-loop ChPT. This suggest
that the inverse-amplitude method is indeed a rather systematic
way of improving ChPT. This conclusion is also supported by a
previous analysis of both the $K_{l4}$ decays and the $\pi\pi /\pi K$
scattering processes \cite{ref:Han95}. The forthcoming DA$\Phi$NE
facility \cite{ref:DAPHNE95} should be an ideal place to test this
conclusion in much greater detail.

\acknowledgments

The author is grateful to A. Miranda and G. C. Oades for useful
discussions and comments, and to J. Bijnens for suggesting this work.
The financial support from The Faculty of Science, Aarhus University
is acknowledged.

\begin{table}
\caption{The low-energy parameters $\langle r^2\rangle$ and $c$ for
the pion and kaon form factors $F_V^{\pi}$, $F_S^{\pi}$, and
$F_V^{K^+}$ obtained from one-loop ChPT (ChPT) and the unitarized
inverse-amplitude (UChPT) method. The experimental information are
from Ref. \protect\cite{ref:NA786} ($\langle r^2\rangle_V^{\pi}$),
Ref. \protect\cite{ref:GM90} ($c_V^{\pi}$), Refs.
\protect\cite{ref:GM90,ref:DGL90}
($\langle r^2\rangle_S^{\pi}$,$c_S^{\pi}$),
and Ref. \protect\cite{ref:Amen86}
($\langle r^2\rangle_V^{K^+}$).}
\label{Table1}
\begin{tabular}{cccc}
&ChPT&UChPT&Experiment\\
\tableline
$\langle r^2\rangle_V^{\pi}$
(${\rm fm}^2$)&0.439$\pm$0.008&0.439$\pm$0.008&0.439$\pm$0.008\\
$c_V^{\pi}$ (${\rm GeV}^{-4}$)&0.7&4.2$\pm$0.1&$4.1-4.3$\\
$\langle r^2\rangle_S^{\pi}$
(${\rm fm}^2$)&0.50$\pm$0.11&0.50$\pm$0.11&$0.57-0.61$\\
$c_S^{\pi}$ (${\rm GeV}^{-4}$)&6.2&10.7$\pm$2.0&$10.0-10.9$\\
$\langle r^2\rangle_V^{K^+}$(${\rm fm}^2$)
&0.40$\pm$0.01&0.40$\pm$0.01&0.34$\pm$0.05\\
\end{tabular}
\end{table}

\begin{table}
\caption{The low-energy parameters $\lambda_+$ and $\lambda_0$ for the
$K_{l3}$ form factors $f_+^{K\pi}$ and $f_0^{K\pi}$ obtained from
one-loop ChPT (ChPT) and the unitarized inverse-amplitude (UChPT)
method. The decay associated with the given values of $\lambda$
is given in parentheses. The experimental data are
from Ref. \protect\cite{ref:PDG94} ($\lambda_+$) and Ref.
\protect\cite{ref:Don74} ($\lambda_0$).}
\label{Table2}
\begin{tabular}{cccc}
&ChPT&UChPT&Experiment\\
\tableline
$\lambda_+$ ($K_{e3}^+$)
&0.0303$\pm$0.0007&0.0289$\pm$0.0006&0.0286$\pm$0.0022\\
$\lambda_+$ ($K_{e3}^0$)
&0.0323$\pm$0.0007&0.0309$\pm$0.0007&0.0300$\pm$0.0016\\
$\lambda_0$ ($K_{\mu 3}^0$)
&0.0176$\pm$0.0013&0.0168$\pm$0.0012&0.019$\pm$0.004\\
\end{tabular}
\end{table}

\pagebreak

\pagebreak

\pagebreak

\pagebreak

\begin{references}
\bibitem{ref:We79} S. Weinberg, Physica {\bf 96A}, 327 (1979).
\bibitem{ref:GL84} J. Gasser and H. Leutwyler, Ann. Phys. (N.Y.) {\bf
158}, 142 (1984); Nucl. Phys. {\bf B250}, 465 (1985).
\bibitem{ref:BCG94} J. Bijnens, G. Colangelo, and J. Gasser, Nucl.
Phys. {\bf B427}, 427 (1994).
\bibitem{ref:BGS94} S. Bellucci, J. Gasser, and M. E. Sainio, Nucl.
Phys. {\bf B423}, 80 (1994); {\bf B431}, 413(E) (1994);
J. Bijnens {\em et al.}, Phys. Lett. B {\bf 374}, 210 (1996).
\bibitem{ref:Tru88} T. N. Truong, Phys. Rev. Lett. {\bf 61}, 2526
(1988).
\bibitem{ref:DHT90} A. Dobado, M. J. Herrero, and T. N. Truong, Phys.
Lett. B {\bf 235}, 134 (1990);
T. N. Truong, Phys. Rev. Lett. {\bf 67}, 2260 (1991);
A. Dobado and J. R. Pel\'{a}ez, Phys. Rev. D {\bf 47}, 4883 (1993); Z.
Phys. C {\bf 57}, 501 (1993); ``The inverse amplitude method in Chiral
Perturbation Theory,'' Report No. hep-ph/9604416.
\bibitem{ref:GM90} J. Gasser and U. G. Meissner, Nucl. Phys. {\bf
B357}, 90 (1991).
\bibitem{ref:BT95} L. Beldjoudi and T. N. Truong, Phys. Lett. B {\bf
351}, 357 (1995).
\bibitem{ref:GL85} J. Gasser and H. Leutwyler, Nucl. Phys. {\bf B250},
517 (1985).
\bibitem{ref:NA786} NA7 Collaboration, S. R. Amendolia {\em et al.},
Nucl. Phys. {\bf B277}, 168 (1986).
\bibitem{ref:PDG94} Particle Data Group, L. Montanet {\em et al.},
Phys. Rev. D {\bf 50}, 1173 (1994).
\bibitem{ref:DGL90} J. F. Donoghue, J. Gasser, and H. Leutwyler, Nucl.
Phys. {\bf B343}, 341 (1990).
\bibitem{ref:Amen86} S. R. Amendolia {\em et al.}, Phys. Lett. B {\bf
178}, 435 (1986).
\bibitem{ref:Bar85} L. M. Barkov {\em et al.}, Nucl. Phys. {\bf B256},
365 (1985).
\bibitem{ref:GS68} G. J. Gounaris and J. J. Sakurai, Phys. Rev. Lett.
{\bf 21}, 244 (1968).
\bibitem{ref:BC88} J. Bijnens and F. Cornet, Nucl. Phys. {\bf B296},
557 (1988).
\bibitem{ref:Amen84} S. R. Amendolia {\em et al.}, Phys. Lett. {\bf
138B}, 454 (1984).
\bibitem{ref:Ani83} G. V. Anikin {\em et al.}, Report No. INP 83-85
(unpublished).
\bibitem{ref:Vas81} I. B. Vasserman {\em et al.}, Yad. Fiz. {\bf 33},
709 (1981) [Sov. J. Nucl. Phys. {\bf 33}, 368 (1981)].
\bibitem{ref:Quen78} A. Quenzer {\em et al.}, Phys. Lett. {\bf 76B},
512 (1978).
\bibitem{ref:Kur84} L. M. Kurdadze {\em et al.}, Yad. Fiz. {\bf 40},
451 (1984)[Sov. J. Nucl. Phys. {\bf 40}, 286 (1984)].
\bibitem{ref:Dal80} E. B. Dally {\em et al.}, Phys. Rev. Lett. {\bf
45}, 232 (1980).
\bibitem{ref:Don74} G. Donaldson {\em et al.}, Phys. Rev. D {\bf 9},
2960 (1974).
\bibitem{ref:Bir81} V. K. Birulev {\em et al.}, Nucl. Phys. {\bf
B182}, 1 (1981).
\bibitem{ref:DAPHNE95} {\em The Second DA$\Phi$NE Physics
Handbook}, edited by L. Maiani, G. Pancheri, and N. Paver
(INFN, Frascati, 1995).
\bibitem{ref:Aki91} S. A. Akimenko {\em et al.}, Phys. Lett. B {\bf
259}, 225 (1991).
\bibitem{ref:Han95} T. Hannah, Phys. Rev. D {\bf 51}, 103 (1995);
{\bf 52}, 4971 (1995).
\end{references}
\end{document}